\documentclass[preprint2]{aastex}
\usepackage{epsfig}
\usepackage{hyperref}
\usepackage{amsmath}
\usepackage{amsfonts}
\usepackage{array}
\begin{document}

\newcolumntype{C}[1]{>{\centering}m{#1}}

\title{$\chi^2$ Discriminators for Transiting Planet Detection in \textit{Kepler} Data}

\author{Shawn Seader\altaffilmark{1}, Peter Tenenbaum\altaffilmark{2}, Jon M. Jenkins\altaffilmark{3}, and Christopher J. Burke\altaffilmark{4}}
\affil{SETI Institute, NASA Ames Research Center, Moffett Field, CA 94035}
\altaffiltext{1}{shawn.seader@nasa.gov}
\altaffiltext{2}{peter.tenenbaum@nasa.gov}
\altaffiltext{3}{jon.jenkins@nasa.gov}
\altaffiltext{4}{christopher.j.burke@nasa.gov}

\begin{abstract}
The \textit{Kepler} spacecraft observes a host of target stars to detect transiting planets.  Requiring a $7.1$ sigma detection in twelve quarters of data yields over $100,000$ detections, many of which are false alarms.  After a second cut is made on a robust detection statistic \cite{Tenenbaum:2012ft}, some $50,000$ or more targets still remain.  These false alarms waste resources as they propagate through the remainder of the software pipeline and so a method to discriminate against them is crucial in maintaining the desired sensitivity to true events.  This paper describes a $\chi^2$ test which represents a novel application of the formalism developed by Allen \cite{Allen:2004gu} for false alarm mitigation in searches for gravitational waves.  Using this technique, the false alarm rate can be lowered to $\sim5\%$.
\end{abstract}


\section{Introduction}
\label{s:intro}
The \textit{Kepler} spacecraft continuously observes more than $150,000$ target stars in a 115 square-degree field of view to discover Earth-like planets transiting Sun-like stars through analysis of photometric data \cite{Borucki:2010zz, Koch:2010vj}.  The \textit{Kepler} spacecraft collects photometric data for each target star which is compressed and stored on-board to be downlinked at monthly intervals.  The \textit{Kepler} Science Operations Center at NASA Ames Research Center processes the data with the Science Processing Pipeline, which is composed of several modules including the Transiting Planet Search (TPS) \cite{Jenkins:2010vf}.  To search for transit signatures, TPS employs a bank of wavelet-based matched filters that form a grid on a three dimensional parameter space of transit duration, period, and epoch \cite{Jenkins, Jenkins2}.  Owing to non-stationary and non-Gaussian noise, uncorrected systematics, and poorly mitigated noise events of either astrophysical or non-astrophysical nature, large spurious Threshold Crossing Events (TCEs) can be produced by the matched filtering performed in TPS.

The optimal linear filter for a deterministic signal buried in Gaussian noise is a simple matched filter \cite{Kay}.  The output of a matched filter can be large, both, when a true signal is contained in the data and when the noise mimmicks the true signal closely enough.  In a perfect detection scenario the noise would be stationary and Gassian, furnishing static false alarm and detection probabilities.  Under this scenario, one can easily apply the Neyman-Pearson criterion to set a detection threshold that maximizes the detection probability for a given, tolerable, false alarm rate.  In reality however, a plethora of outside influences can contribute non-stationary and non-Gaussian noise to the data stream.  An effort can and should be made to understand as many of these periphereal noise sources as possible and remove their effect through model fitting.  Imperfections in the removal of known noise sources and other spurious noise events may then however necessitate the formulation of a discrimination strategy for the remaining false alarms that contribute to a false alarm rate that is above what is expected from consideration of Gaussian statistics alone.  The chosen discrimination method should have as a requirement to preserve the existing detection probability while simultaneously reducing the false alarm rate.  True positives should easily pass the test, while false positives should be mitigated.

A clever approach to this problem was developed by Allen \cite{Allen:2004gu} for use in gravity wave searches in LIGO data.  In that work, the chief astrophysical source of interest was the inspiraling compact binary system.  The signal, or chirp, from such a system is a sinusoid with both a frequency and amplitude that diverge as the stars approach the merger phase of the inspiral event.  The $\chi^2$ discriminator is essentially built by breaking up the frequency band of the detector into chunks and for each chunk comparing the expected response, given the template waveform, to the actual response.  Allen \cite{Allen:2004gu} shows that the expectation value of the $\chi^2$ statistic is independent of whether or not a signal is present in the data.  This property makes it a good discriminator since its value deviates from the expectation value only when the noise deviates from its assumed Gaussianity and stationarity.  So it provides a method for targeting contributions from non-Gaussian tails that end up furnishing TCEs.  Allen's work goes on to show that under the assumptions of both stationary and Gaussian noise, the $\chi^2$ statistic can be proven to have a $\chi^2$ probability density function.  The paper also shows that when there is mismatch between the true signal and the template, the probability density function of the $\chi^2$ statistic then becomes a non-central $\chi^2$ distribution whose non-centrality parameter depends on both the signal-template mismatch and the square of the expected SNR. 

Allen's method was originally formulated for use with broad-band deterministic signals, with unkown signal parameters, and detectors with potentially non-ideal observing noise.  In the case of planet detection, the problem is similar.  The waveforms used by TPS depend on a set of three parameters, namely, epoch, period, and transit duration ($\{t_0, T, d\}$ respectively).  Planetary transits produce periodic depressions in the light curves of their host stars.  The exact shape of the depression depends strongly on both stellar and planetary parameters.  TPS avoids this complication by simply using pulse trains of square waves, which means that the templates and true signal are inherently mismatched.  The degree of mismatch is furthered by the fact that only a discrete set of points in the three dimensional parameter space is searched over.  The goal of TPS is to simply identify those targets that potentially have something interesting and should be followed up with the final Pipeline module, Data Validation (DV).  DV then does a much more thorough job of fitting an astrophysical transit model to determine the validity of the TCE.

Although the signal detection problems differ between the gravity wave detection, for which the $\chi^2$ formalism was originally developed, and the transiting planet detection, the basic ideas can still be applied.  Rather than breaking up the detector's bandwidth into different channels, we instead analyze the components of our test statistics in multiple ways that are completely analogous with the $\chi^2$ formalism developed by Allen.  Many of the results can then be directly applied.

This paper is organized as follows.  Section~\ref{s:detection} reproduces the detection theory employed by TPS for completeness.  Section~\ref{s:chisquares} then applies the $\chi^2$ formalism to produce a discriminator for TPS and discusses its expected distribution.  Section~\ref{s:otherchis} discusses some of the other possible versions of the $\chi^2$ discriminator.  Section~\ref{s:signaltemplatemismatch} shows what the effect of a mismatch between the signal and template has on the $\chi^2$ statistic. In this section the true mismatch is estimated by a Monte Carlo study which also proves to be useful for generating an astrophysically motivated set of templates for TPS.  Section~\ref{s:thresholds} shows how TPS uses the discriminators for thresholding purposes now and and how it might use them in the future. Section~\ref{s:examples} gives some examples that illustrate how the $\chi^2$ discriminators work and what their strengths and weaknesses are.  Section~\ref{s:results} gives some results based on analysis of twelve quarters of \textit{Kepler} data.  Then Section~\ref{s:conclusion} gives a conclusion which summarizes the main results, gives prospects for future work, followed by acknowledgments.  

\section{Detection Theory}
\label{s:detection}
The data input to TPS are discrete, contiguous, flux fraction time series that have been corrected for systematics and had some other more localized noise artifacts removed such as sudden pixel sensitivity dropouts, cosmic rays, and thermal transients.  For a discussion of how the data are prepared for the search in TPS see \cite{Tenenbaum:2012ft}.  Let $x(n)$ be this discrete, contiguous, flux fraction time series, where $n \in [1,..., N]$.  Under the null hypothesis, $H0$, there is no transit signal present and we have only noise $w(n)$ which we assume to be zero-mean, White Gaussian Noise (WGN), with variance $\sigma^2$.  Under the alternative hypothesis, $H1$, there is a transit pulse signal $s(n)$ present in the data (for simplicity, assume for now a single pulse is present rather than a pulse train).  We then have:
\begin{equation}
\label{eq1}
\begin{split}
&H0: x(n) = w(n)\\
&H1: x(n) = w(n) + s(n).
\end{split}
\end{equation}
An often used result from detection theory is that for a known signal, the optimal detection statistic is a simple matched filter of the form
\begin{equation}
\label{eq2}
z=\frac{\textbf{x} \cdot \textbf{s}}{ \sigma \sqrt{\textbf{s} \cdot \textbf{s}} },
\end{equation}
where we have used bold-faced type to denote vector quantities and $\sigma$ is the standard deviation of $w(n)$.  Since $z$ is a linear combination of Gaussian random variables, it too is a Gaussian random variable.  The mean and variance of $z$ under the two hypotheses therefore completely characterize the detection problem.  It is straightforward to show that under both hypotheses, the variance of $z$ is unity, while the mean under $H0$ is zero and the mean under $H1$ is the SNR of the signal:
\begin{eqnarray}
\nonumber
{\langle z \rangle}_{H0} & = & {\bar{z}}_0 \\
                        & = & 0,
\end{eqnarray}
\begin{eqnarray}
\nonumber
{\langle z \rangle}_{H1} & = & {\bar{z}}_1 \\
                        & = & \frac{\sqrt{\textbf{s} \cdot \textbf{s}}}{ \sigma},
\end{eqnarray}
\begin{eqnarray}
\nonumber
{\langle (z - {\bar{z}}_0)^2 \rangle}_{H0} & = & {\langle (z - {\bar{z}}_1)^2 \rangle}_{H1} \\
                                         & = & 1.
\end{eqnarray}
Now it is easy to see that the probability density functions under the two hypotheses are given by:
\begin{equation}
{p}_{i}(x) = \frac{1}{\sqrt{2\pi}} \exp[\frac{ -(x-{\bar{z}}_i)^2}{2}] , \ \ i = \{0,1\},
\end{equation}
and the corresponding false alarm and detection probabilities as a function of the decision making threshold $\eta$ are:
\begin{equation}
P_{FA} = \int_\eta^\infty p_0(x) \; dx
\end{equation}
\begin{equation}
P_D = \int_\eta^\infty p_1(x) \; dx.
\end{equation}
In this detection scenario there is no knowledge of prior probabilities nor any costs associated with false alarms or false dismissals, so the best strategy to adopt is to apply the Neyman-Pearson criteria.  A maximum tolerable false alarm is chosen which determines the threshold $\eta$ and the corresponding detection probability.  

Since we are observing the light from many target stars with highly varied properties, the noise $w(n)$ is typically not white, but colored.  Colored Gaussian noise can be modeled as the result of filtering white Gaussian noise through a linear but possibly time-varying filter \cite{Hayes}.  The noise will generally then possess an auto-correlation matrix, $\textbf{R}$, with non-zero off-diagonal components.  The optimal detector is still the simple matched filter but it must account for the off-diagonal components:
\begin{equation}
\label{autocorr}
z = \frac{\textbf{x}^T\textbf{R}^{-1}\textbf{s}}{\sqrt{\textbf{s}^T\textbf{R}^{-1}\textbf{s} } }.
\end{equation}
Since $\textbf{R}$ is an autocorrelation matrix of the noise $w(n)$, it is non-singular and symmetric and therefore possesses a square root.  So we can rewrite (\ref{autocorr}) as
\begin{eqnarray}
\nonumber
z &=& \frac{(\textbf{R}^{-1/2}\textbf{x})^T (\textbf{R}^{-1/2}\textbf{s})}{\sqrt{(\textbf{R}^{-1/2}\textbf{s})^T(\textbf{R}^{-1/2}\textbf{s})}} \\
  &=& \frac{\tilde{\textbf{x}} \cdot \tilde{\textbf{s}}}{\sqrt{\tilde{\textbf{s}} \cdot \tilde{\textbf{s}}}},
\end{eqnarray}
where $\tilde{\textbf{x}} = \textbf{R}^{-1/2}\textbf{x}$ and $\tilde{\textbf{s}} = \textbf{R}^{-1/2}\textbf{s}$ are whitened versions of the data and signal vectors.  Since $R$ is typically not available, one is faced with the difficult task of designing an appropriate whitening filter to construct the test statistic in the presence of colored noise.  Since the stellar irradiance can clearly exhibit (colored) nonstationary behaviour, a wavelet-based adaptive matched filter that constructs a time-varying whitener is employed \cite{Jenkins,Jenkins2}.  This detector performs a joint time-frequency decomposition of the data to estimate the properties of the noise as a function of time, then applies a matched filter to the whitened data in the wavelet domain, taking into account the effect of the whitener on the shape of the transit pulse.

The wavelet based approach employed by TPS uses an Over-complete (discrete-time) Wavelet Transform (OWT) of the data and template with Debauchies' 12-tap wavelets \cite{debauchies}.  The wavelet domain is a natural choice for designing time-varying filters since it \textit{is} a joint time-frequency representation of the transformed data.  The wavelet-based matched filter uses an octave-band filter bank to separate input flux time series into different band passes to estimate the noise Power Spectral Density (PSD) as a function of time. For details of the filter bank implementation see \cite{Jenkins, Jenkins2}.  For our purposes here, define the OWT of $x(n)$ as
\begin{equation}
\label{waveletTransform}
\mathbb{W}\{x(n)\} = \{ x_1(n), x_2(n), ..., x_M(n) \},
\end{equation}
where
\begin{equation}
x_i(n) = h_i(n) \ast x(n) \ , \ i = 1, 2, ..., M, 
\end{equation}
`$\ast$' denotes convolution, and the $h_i(n)$ are the impulse responses of the filters in the filter bank implementation of the wavelet expansion with corresponding frequency responses $H_i(\omega)$.  The filter $H_1$ is a high-pass filter that passes frequency content from half the Nyquist frequency to Nyquist ($[f_{Nyquist}/2, f_{Nyquist}]$).  The next filter, $H_2$, then passes frequency content in the interval $[f_{Nyquist}/4, f_{Nyquist}/2]$. Each subsequent filter passes content in the next lower bandpass until the final filter, $H_M$, passes the lowest bandpass on down to DC.  The time-varying channel variance, $\hat{\sigma}_i^2$, in each channel $i$ of the filter bank is estimated by a moving circular median absolute deviation (MAD) with an analysis window chosen to be significantly longer than the duration of transit pulse.  

To obtain the wavelet-based expression for the matched filter, we need to be able to express the dot product in the wavelet domain.  For an overcomplete, dyadic, wavelet expansion, the dot product can be expressed as:
\begin{equation}
\label{waveletdot}
\textbf{x} \cdot \textbf{y} = \sum_{i=1}^M 2^{-\min(i,M-1)} \textbf{x}_i \cdot \textbf{y}_i,
\end{equation}
where $\textbf{x}$ and $\textbf{y}$ are time series, and $\textbf{x}_i$ and $\textbf{y}_i$ are the wavelet components which are also time series \cite{Vetterli}.  The restriction on the power of two in (\ref{waveletdot}) is necessary because the last two channels of the OWT have the same bandwidth.  The detection statistic is now computed by multiplying the whitened wavelet coefficients of the data by the whitened wavelet coefficients of the transit pulse and employing the dot product relation:
\begin{eqnarray}
\nonumber
z & = & \frac{\tilde{\textbf{x}} \cdot \tilde{\textbf{s}}}{\sqrt{\tilde{\textbf{s}} \cdot \tilde{\textbf{s}}}} \\
\nonumber
  & = & \frac{\sum_{i=1}^M 2^{-\min(i,M-1)} \sum_{n=1}^N x_i(n) s_i(n) \slash \hat{\sigma}_i^2(n)}{\sqrt{\sum_{i=1}^M 2^{-\min(i,M-1)} \sum_{n=1}^N s_i^2(n) \slash \hat{\sigma}_i^2(n)}}. \\
\label{SESOrig}
\end{eqnarray}  

This equation (\ref{SESOrig}) is the Single Event Statistic (SES).  It allows us to compute the detection statistic for a transit pulse $\textbf{s}$ of a specific duration, where the transit is centered at a single point in time $n$. The $\hat{\sigma}_i^2(n)$ are the whitening coefficients which are variance estimates for each time point and for each wavelet scale $i$.  These are estimated by a moving circular Median Absolute Deviation (MAD) at each scale with a base window size that is thirty times longer than the pulse duration.

At the outset of our analysis, the timing of a real transit signal, namely its epoch and period, is unknown.  Therefore, it would be useful to generate a time series of detection statistics, $z(n)$, which represents the detection statistic at sample $n$ for a single transit pulse which is centered at $n$.  This can be done via $N$ repeated invocations of equation (\ref{SESOrig}), in which we recenter the transit pulse at all possible times ${n}$ sequentially.  Although effective, this method of computing $z(n)$ is cumbersome, requiring that the wavelet transform of equation (\ref{waveletTransform}) and the dot product in equation (\ref{SESOrig}) be performed for each of the $N$ unique transit pulse models $\textbf{s}$.  An equivalent result can be achieved more efficiently by recognizing that the mass of invocations of equation (\ref{SESOrig}) is equivalent to computing a cross correlation, in which the array of different $\textbf{s}$ vectors is replaced by a single one, representing a single transit of unit depth centered at $n=1$.  
     
To compute the detection statistic $z$ for a given transit pulse centered at all possible time steps, we can simply doubly whiten $\mathbb{W}\{x(n)\}$ (i.e. divide $x_i(n)$ point-wise by $\hat{\sigma}_i^2(n)$), correlate the results with $\mathbb{W}\{s(n)\}$, and apply the dot product relation, performing the analogous operations for the denominator, noting that $\hat{\sigma}_i^{-2}(n)$ is itself a time series:
\begin{eqnarray}
\label{SES}
\nonumber
z(n) & = & \frac{\mathbb{N}(n)}{\sqrt{\mathbb{D}(n)}} \\
\nonumber
     & = & \frac{\sum_{i=1}^M 2^{-\min(i,M-1)} [x_i(n) \slash \hat{\sigma}_i^2(n)] \ast s_i(-n)}{\sqrt{\sum_{i=1}^M 2^{-\min(i,M-1)} \hat{\sigma}_i^{-2}(n) \ast s_i^2(-n) }}. \\
\end{eqnarray}
Note that the `$-$' in $s_i(-n)$ indicates time reversal.  The $\mathbb{N}(n)$ and $\mathbb{D}(n)$ are introduced for convenience later on.

The quantity $z(n)$ in equation (\ref{SES}) is referred to as the Single Event Statistic (SES) time series.  Note that the quantity $\sqrt{\mathbb{D}(n)}$ is the epected SNR of the template in the data as a function of time, or the expected SNR of a true signal that matches the shape and amplitude of the template exactly.  To make explicit the dependence on the signal amplitude, $\mathcal{A}$, under $H1$ let:
\begin{equation}
\textbf{x} = \textbf{w} + \mathcal{A} \textbf{s}.
\end{equation}
The relevant statistical quantities are then given by:
\begin{equation}
\label{expectedSES}
{\langle z(n) \rangle} = \mathcal{A} \sqrt{\mathbb{D}(n)},
\end{equation}
\begin{equation}
\label{SESVar}
{\langle z^2(n) \rangle} = 1 + \mathcal{A}^2 \mathbb{D}(n),
\end{equation}
where, under $H0$, we can simply let $\mathcal{A} \rightarrow 0$.  So under either hypothesis, the SES has unit variance.  

  Up until now we have only explicitly used one of the signal parameters, namely, the transit duration $d$, which is built into the template.  The TPS module currently searches over 14 trial transit durations logarithmically spaced between 1.5 hours up to 15 hours.  To perform the search over the remaining two parameters, period and phase ($T$ and $t_0$ respectively), we must lay down a grid in the parameter space that balances the need to preserve sensitivity to the astrophysically interesting parameter space with the need for computational tractability.  The sensitivity requirements dictate the spacing on the parameter space.  For a discussion of this see \cite{Jenkins2}.  To perform the search over period and phase, the single event statistics must be folded using each discrete point in parameter space.  In practice however, to reduce computation, a more sophisticated folding routine is employed that ensures all interesting portions of parameter space get searched while uninteresting portions are skipped over on a target-by-target basis \cite{Tenenbaum:2012ft}.  A description of how to compute the Multiple Event Statistic (MES), $Z(t_0,T,d)$, is given here.

Choosing a particular point in the $\{T,t_0\}$ space selects out a set, $\mathcal{S}$, of $P$ samples, one for each transit, that start with the sample corresponding to the epoch $t_0$ and are spaced $T$ samples apart.  These samples form a subset of $\{n\}$, $\mathcal{S} = \{t_0,t_0+T,...,t_0+(P-1)T\}$.  The MES is then constructed as:
\begin{equation}
\label{MES}
Z = \sum_{i \in \mathcal{S}} \mathbb{N}(i) \slash \sqrt{ \sum_{i \in \mathcal{S}} \mathbb{D}(i) }.
\end{equation}

\section{\texorpdfstring{$\chi^2$}{x2} Discriminator}
\label{s:chisquares}
This section will derive a version of the $\chi^2$ statistic that has been found to be useful in the analysis of \textit{Kepler} data.  A brief summary of the formalism developed by Allen is first in order. 

The basic idea behind the construction of the test statistic is to break up the matched filter output into several contributions and compare each contribution with what is expected.  What follows in this paragraph is taken from \cite{Allen:2004gu} for completeness.  First, the detector output $z$ is broken up into $p$ chunks.  Mathematically we have,
\begin{equation}
z = \sum_{j=1}^p z_j,
\end{equation}
where the $z_j$ are additive chunks of the filter output that when added together reproduce exactly the output value of the filter.  These are the actual contributions to the filter output.  Next consider the $p$ quantities defined by
\begin{equation}
\Delta z_j \equiv z_j - q_jz,
\end{equation} 
where
\begin{equation}
\sum_{j=1}^p q_j = 1,
\end{equation}
and the $q_j$ are the expected fractional contribution to $z$ from the $j$'th chunk.  The $\Delta z_j$ are then the set of differences between the $p$ actual contributions and expected contributions. By definition, the $\Delta z_j$'s sum to zero
\begin{equation}
\sum_{j=1}^p \Delta z_j = 0,
\end{equation} 
and their expectation values vanish
\begin{equation}
\langle \Delta z_j \rangle = 0.
\end{equation}
The $\chi^2$ statistic is then defined as
\begin{equation}
\chi^2 = \sum_{j=1}^p  (\Delta z_j)^2 \slash q_j.
\end{equation}
Note that with some basic assumptions on the detector noise, the expectation value of this statistic is independent of whether or not a signal is present in the data, thereby making this an ideal discriminator for noise events.  The noise considered here is assumed to meet the following criteria:
\begin{equation}
{\langle \tilde{w}(n) \rangle} = 0,
\end{equation}
\begin{equation}
{\langle \tilde{w}(n) \tilde{w}(m) \rangle} = \delta(n-m),
\end{equation}
where $\delta(n)$ is the Dirac delta function.  So we assume the whitened noise has zero mean, unit variance, and is uncorrelated.  

This concept can now be applied to the MES given in (\ref{MES}) by breaking it up into a set of $P$ contributions, where again, $P$ is the number of transits.  The MES calculation was done in the wavelet domain to properly whiten the data and templates.  Due to the imperfect localization of the filters in the OWT however, the wavelet components become correlated in an intricate way.  In order to eliminate this correlation effect we can simply apply the inverse OWT on the whitened data and template prior to computing the veto.  This will also make it easier to compute the statistical properties of the quantities of interest as well as shed light on some of the subtle issues surrounding the calculation. The MES can be re-written as:
\begin{eqnarray}
\nonumber
Z &=& \frac{\sum_{j=1}^P \tilde{\textbf{x}} \cdot \tilde{\textbf{s}}_j}{\sqrt{\sum_{j=1}^P \tilde{\textbf{s}}_j \cdot \tilde{\textbf{s}}_j}} \\
  &=& \frac{\sum_{j=1}^P \sum_{n=1}^N \tilde{x}(n) \tilde{s}_j(n)}{\sqrt{\sum_{j=1}^P \sum_{n=1}^N \tilde{s}^2_j(n)}},
\end{eqnarray}
where `$\sim$' denotes a whitened vector and $s_j(n)$ is a template with a transit pulse centered at the time corresponding to transit $j \in \mathcal{S}$.  Now, let
\begin{equation}
z_j = \frac{\sum_{n=1}^N \tilde{x}(n) \tilde{s}_j(n)}{\sqrt{\sum_{k=1}^P \sum_{n=1}^N \tilde{s}^2_k(n)}},
\end{equation}
so clearly we have:
\begin{equation}
Z = \sum_{j=1}^P z_j.
\end{equation}
Similarly, identify $q_j$ as
\begin{equation}
q_j = \frac{\sum_{n=1}^N \tilde{s}^2_j(n)}{\sum_{k=1}^P \sum_{n=1}^N \tilde{s}^2_k(n)},
\end{equation}
where clearly
\begin{equation}
\sum_{j=1}^P q_j = 1.
\end{equation}
The $z_j$ are the temporal contributions to the MES and the $q_j$ are the fractional expected temporal contributions. 
The $\Delta z_j$ and $\chi^2$ statistic can be constructed as:
\begin{equation}
\Delta z_j = z_j - q_j Z
\end{equation}
\begin{equation}
\chi^2 = \sum_{j=1}^P \frac{(\Delta z_j)^2}{q_j}.
\end{equation}

There are three subtle issues involved in this calculation that have been neglected.  The first issue is related to the calculation of the whitening coefficients, or the $\sigma$'s, in (\ref{SESOrig}).  The noise is estimated at each wavelet scale by a moving circular MAD filter.  This method is robust against outliers but if there is a planetary transit signal in the data then it can perturb the whitening coefficients.  The $\chi^2$ calculation requires that the $q_i$ components be explicitly independent of the presence of a transit signal.  Therefore, prior to computing the whitening coefficients, the in-transit samples are gapped and filled using an auto-regressive algorithm to guarantee the necessary signal independence of the whitening coefficients.

To understand the next subtlety associated with the calculation of the veto, consider that under $H1$, the data can in general be written:
\begin{equation}
\tilde{x}(n) = \tilde{w}(n) + \mathcal{A} \tilde{s}(n),
\end{equation}
where $\tilde{s}(n)$ is a transit pulse train rather than a transit centered at some time corresponding to transit $j$ as in the template version of $\tilde{s}_j(n)$.  This difference introduces correlation in the $z_j$'s.  This issue can be handled in a similar manner as the whitening coefficients above.  To get rid of this correlation we simply have to gap and fill all the in-transit samples that are not associated to the transit identified by $j$ prior to computing each $z_j$.  This effectively turns $\tilde{x}(n)$ into $\tilde{x}_j(n)$:
\begin{equation}
\tilde{x}_j(n) = \tilde{w}(n) + \mathcal{A} \tilde{s}_j(n).
\end{equation}

The final subtlety is that after whitening our template in the wavelet domain we have to zero out all the out-of-transit samples since the effect of the transit gets smeared out across more samples even after doing the inverse OWT.  In fact, there would not be any out-of-transit samples with a value of zero in the template without the windowing.  This correction is necessary so that the $z_j$ components achieve the correct statistical properties and so the $\Delta z_j$ have the correct correlation structure.  Note also that this windowing will make $Z \neq MES$ in general.  In what follows it is assumed that these subtleties are being corrected for as stated above.

The relevant statistical properties of the various quantities of interest can now be computed and summarized:
\begin{equation}
\label{statProps}
\begin{split}
&{\langle Z \rangle} = \mathcal{A} \mathcal{D}, \\
&{\langle Z^2 \rangle} = 1 + \mathcal{A}^2 \mathcal{D}^2, \\
&{\langle z_j \rangle} = q_j \mathcal{A} \mathcal{D}, \\
&{\langle z_j^2 \rangle} = q_j + q_j^2 \mathcal{A}^2 \mathcal{D}^2, \\
&{\langle z_jz_k \rangle} = q_j\delta_{jk} + q_jq_k \mathcal{A}^2 \mathcal{D}^2, \\
&{\langle \Delta z_j \rangle} = 0, \\
&{\langle (\Delta z_j)^2 \rangle} = q_j(1-q_j), \\
&{\langle \chi^2 \rangle} = P - 1, \\
&{\langle (\chi^2)^2 \rangle} = P^2 - 1, \\
\end{split}
\end{equation}
where,
\begin{equation}
\mathcal{D} = \sqrt{ \sum_{j=1}^P \sum_{n=1}^N \tilde{s}^2_j(n) }
\end{equation}
the $\delta_{ij}$ is the Kronecker delta, and the statistical properties under $H0$ can be obtained by letting $\mathcal{A} \rightarrow 0$.

\begin{figure}
\begin{center}
\epsfig{file=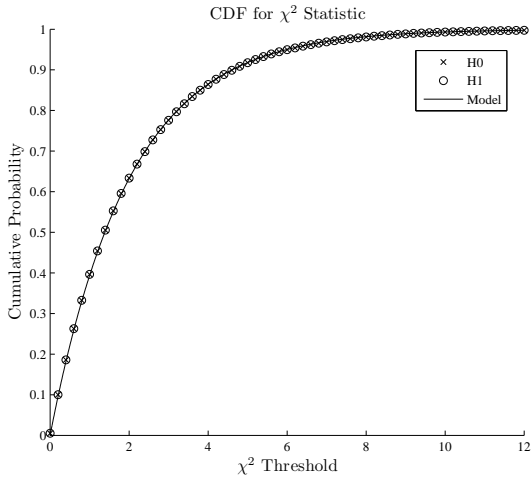,width=3.4in}
\caption{ \label{f:cdfMonteCarlo} Cumulative Distribution of the $\chi^2$ statistic resulting from the 100,000 sample Monte Carlo under H0 and H1 with Gaussian noise.  Time Series of length $1024$ samples were used.  Injected transits had a duration of $12$ samples and were spaced $400$ samples apart.  The model here is a $\chi^2$ cumulative distribution function for a $\chi^2$ with two degrees of freedom.  }
\end{center}
\end{figure}

Using the afforementioned assumptions on the detector noise and also assuming a perfect match between the signal and template, a proof was given in \cite{Allen:2004gu} that this $\chi^2$ statistic is $\chi^2$-distributed with $P-1$ degrees of freedom\footnote{In \cite{Allen:2004gu}, the unequal expected SNR interval case was treated with the additional assumption that the signal and template did not match exactly.  In that case the distribution of $\chi^2$ was proven to be a non-central $\chi^2$ distribution with $P-1$ degrees of freedom and a non-centrality parameter that was proportional to the mismatch and the SNR squared.  Since we have assumed zero mismatch between the signal and template, the non-centrality parameter is zero and therefore the non-central $\chi^2$ distribution becomes the (central) $\chi^2$ distribution.}.  The cumulative probability that $\chi^2 < \chi^2_0$ is given by:
\begin{eqnarray}
\nonumber
P_{\chi^2<\chi^2_0} &=& \int_0^{\frac{\chi^2_0}{2}} \frac{u^{(\frac{P}{2}-\frac{3}{2})} e^{-u}}{\Gamma(\frac{P}{2}-\frac{1}{2})} \ du \\
\label{cumProb}
               &=& \frac{\gamma(\frac{P}{2}-\frac{1}{2}, \frac{\chi^2_0}{2})}{\Gamma(\frac{P}{2}-\frac{1}{2})}
\end{eqnarray}
where $\gamma$ is the incomplete gamma function.  To check for this expected behavior we performed a 100,000 sample Monte Carlo under both $H0$ and $H1$.  To speed up the test we used 1024 sample time series of random white noise generated with different noise seeds.  In the run under $H1$ we then injected three square waves that were 12 samples long and spaced 400 samples apart.  Each injected square wave had a depth of $8/\sqrt{3}$.  The cumulative distribution functions for each hypothesis are shown in Figure~\ref{f:cdfMonteCarlo}.  The curves under both the null and alternative hypotheses match with the theoretical model extremely well for this Gaussian noise case.

Note that this version of the $\chi^2$ discriminator has been referenced in Appendix A of \cite{Tenenbaum:2012ft}.  There, it is referred to as $\chi^2_{(2)}$.  The results presented in that work however use an older version that had not been fixed to account for the subtleties mentioned above.  The version presented in that paper was also cast in the wavelet domain, which is flawed due to the correlation introduced as mentioned above.  Results given here in section~\ref{s:results} are obtained with an updated codebase that takes into account all the subtleties and uses this re-cast time domain version.  To allow for easier comparison, we will refer to this version of the $\chi^2$ in subsequent sections of this paper as $\chi^2_{(2)}$.

\section{Other \texorpdfstring{$\chi^2$}{x2} Tests}
\label{s:otherchis}
The $\chi^2$ statistic presented in section~\ref{s:chisquares} involves splitting up the MES into its temporal contributions.  In addition to this version, several other versions have also been explored that break up the detector output in other ways.  One such alternate method  analyzes the wavelet contributions to the SES.  Starting with (\ref{SES}), we can identify the $z_i$ and $q_i$ wavelet components as:
\begin{eqnarray}
\nonumber
z_i(n) & = & \frac{\mathbb{N}_i(n)}{\sqrt{\mathbb{D}(n)}} \\
\nonumber
       & = & \frac{2^{-\min(i,M-1)} [x_i(n) \slash \hat{\sigma}_i^2(n)] \ast s_i(-n)}{\sqrt{\sum_{k=1}^M 2^{-\min(k,M-1)} \hat{\sigma}_k^{-2}(n) \ast s_k^2(-n) }} \\
 &
\end{eqnarray}
and
\begin{eqnarray}
\nonumber
q_i(n) & = & \frac{\mathbb{D}_i(n)}{\mathbb{D}(n)} \\ 
\nonumber
       & = & \frac{2^{-\min(i,M-1)} \hat{\sigma}_i^{-2}(n) \ast s_i^2(-n)}{\sum_{k=1}^M 2^{-\min(k,M-1)} \hat{\sigma}_k^{-2}(n) \ast s_k^2(-n) }, \\
&
\end{eqnarray}
where now the $z_i(n)$ are the actual contributions the the SES time series from the $i$'th wavelet component and $q_i(n)$ are the corresponding expected contributions.  The $\mathbb{N}_i(n)$ and $\mathbb{D}_i(n)$ are the wavelet components of the previously defined $\mathbb{N}(n)$ and $\mathbb{D}(n)$.  Now the $\chi^2$ statistic can be formed:
\begin{equation}
\Delta z_i(n) = z_i(n) - q_i(n)z(n)
\end{equation}
\begin{equation}
\chi^2(n) = \sum_{i=1}^M \frac{[\Delta z_i(n) ]^2}{q_i(n)},
\end{equation}
where $M$ is the number of wavelet scales and is determined by the number of data samples $N$ and the length of the mother wavelet filter chosen to implement the filter bank.  This statistic would be $\chi^2$ distributed with $M-1$ degrees of freedom if there was no overlap between the wavelet scales. Since there is overlap however, using this statistic for vetoing purposes can be dangerous because the overlap is difficult to model.   We have a value for this statistic at each $j \in \mathcal{S}$, so we can form a coherent statistic by adding up the $P$ points that contribute to the MES.  This gives
\begin{eqnarray}
\nonumber
\chi_{(1)}^2 & = & \sum_{j \in \mathcal{S}} \chi^2(j) \\
\nonumber
           & = &  \sum_{j \in \mathcal{S}} \sum_{i=1}^M \frac{[\Delta z_i(j)]^2}{q_i(j)}. \\
\end{eqnarray}
The degrees of freedom in the perfect case with no overlap between the wavelet scales would then become $P(M-1)$.  In reality however, the overlap lowers the degrees of freedom and alters the correlation structure of the $\Delta z_i$'s thereby skewing the distribution form.  This version of the statistic has been referenced in Appendix A of \cite{Tenenbaum:2012ft}. There it was referred to as $\chi_{(1)}^2$ as it is now being refferred to in this paper.  There is no difference in the way this version was formulated in \cite{Tenenbaum:2012ft}.

In a similar way, we could also analyze the wavelet contributions to the MES.  
\begin{equation}
\label{zNew}
Z_i = \frac{\sum_{j \in \mathcal{S}} \mathbb{N}_i(j)}{\sqrt{\sum_{j \in \mathcal{S}} \sum_{k=1}^M \mathbb{D}_k(j)}}
\end{equation}
\begin{equation}
Q_i = \frac{\sum_{j \in \mathcal{S}} \mathbb{D}_i(j)}{\sum_{j \in \mathcal{S}} \sum_{k=1}^M \mathbb{D}_k(j)},
\end{equation}
where $Z_i$ are the actual wavelet contributions to the MES and the $Q_i$ are the expected wavelet contributions.  Now, $\chi^2$ can be constructed:
\begin{equation}
\Delta Z_i = Z_i - Q_i Z
\end{equation}
\begin{equation}
\chi^2 = \sum_{i=1}^M \frac{\Delta Z_i^2}{Q_i},
\end{equation}
where now $Z$ is the sum over $i$ of $Z_i$ in equation (\ref{zNew}).  Again however, this statistic is not $\chi^2$-distributed with $M-1$ degrees of freedom due to the overlap between the wavelet scales. This statistic has not proven to be very useful, so it is being omitted from the numbering scheme. 

A version more akin to a classical $\chi^2$ statistic can be formulated by simply comparing each SES that contributes to the MES with what we expect.  The observed value for each SES is given by equation (\ref{SESOrig}), where we can make it more explicit that the transit pulse is centered at some time $n=j$ where $j \in \mathcal{S}$.  To avoid the subtle correlation issues that arise in the wavelet domain we also use whitened time domain vectors:
\begin{equation}
z_j = \frac{\tilde{\textbf{x}} \cdot \tilde{\textbf{s}}_j}{\sqrt{\tilde{\textbf{s}}_j \cdot \tilde{\textbf{s}}_j}}. \\
\end{equation}  
The expected value is given by equation (\ref{expectedSES}), again with the pulse centered at a specific $n=j$:
\begin{equation}
{\langle z_j \rangle}  =  \mathcal{A} \sqrt{\tilde{\textbf{s}}_j \cdot \tilde{\textbf{s}}_j}, \\
\end{equation} 
where A is the signal depth which can be estimated by robustly fitting the full whitened trial pulse train to the whitened data.  With these components we can construct the classical $\chi^2$ as:
\begin{equation}
\chi_{(3)}^2 =  \sum_{j=1}^P \frac{(z_j - {\langle z_j \rangle})^2}{\langle z_j \rangle}.
\end{equation}

Other versions can be formulated that analyze the contributions to the SES and MES from each in-transit sample as well.  These tend not to work well however since the number of degrees of freedom can become very high, thereby diluting the effect of any glitch that might be causing problems for the detector.  Of course, still more versions can be formulated by mixing these up in all possible permutations, for example, analyzing the wavelet contributions to the in-transit samples. 

\section{Effect of Signal/Template Mismatch}
\label{s:signaltemplatemismatch} 

   Up to this point it has been assumed that the true astrophysical signal matches exactly with one of the templates in the template bank.  In TPS currently however, this is never the case.  The Transiting Planet Search currently uses as its model a square wave pulse train that is parameterized by three parameters, namely, epoch or phase, period, and duration ($\{t_0, T, d\}$ respectively).  A discrete grid of templates on this three dimensional parameter space is used for the search, so even if a true astrophysical signal were square wave in shape, there would be mismatch induced by the discrete nature of the search grid.  In transit duration, TPS uses a logarithmically spaced set of 14 different values from 1.5 to 15 hours.  For period and epoch, there is a minimum correlation requirement of 0.9 between neighboring templates that is used to determine the spacing.  Periods are searched in the range from a half a day out to half the length of the data since we currently require 3 transits for a detection.  The epoch is then searched over $[0,T]$.  The mismatch induced by using this discrete three dimensional grid then adds to whatever inherent shape mismatch there is between the square wave pulse and the true astrophysical signal.  The size of the total mismatch drives down the signal-to-noise ratio (SNR).  This gives rise to a sort of balancing act between detection efficiency and computational time or tractability of the search.  For a discussion of the correlation coefficient and how it relates to the match between neighboring templates see \cite{Jenkins2} or \cite{1996Icar}.

   When a target is identified by TPS as containing a threshold crossing event it gets sent on to the Data Validation (DV) portion of data processing pipeline.  Here, a model fitting algorithm fits an astrophysical model to the light curve.  The astrophysical model employed is the geometric transit model of Mandel and Agol \cite{Mandel:2002vx}.  The limb darkening is taken into account using the non-linear limb darkening model of Claret \cite{Claret} which depends on some stellar parameters for the target star, namely its effective temperature, metallicity, and surface gravity.  This level of complexity is required to produce the correct transit shape, which is not a square wave. In what follows we will assume that the true signal is perfectly described by this transit model pulse train that DV produces. 

Here we will consider only how the mismatch between signal and template can affect the distribution of $\chi^2_{(2)}$ of section~\ref{s:chisquares}.  Recall that $\chi^2_{(2)}$ examines the temporal contributions to the MES.  Effectively then, each piece of the statistic is independent of the period since they are all treated individually.  The mismatch in period simply causes the timing offset for a given transit to be a function of the transit time.  So the period mismatch can just be lumped together with the epoch mismatch when we consider how much an individual transit can potentially be mismatched in time from the real transit.  

   To begin, let $\tilde{s}$ and $\tilde{s}'$ denote the whitened, windowed, template and true astrophysical transit respectively, centered at some arbitrary time. Now under $H1$ the true signal is present in the data:
\begin{equation}
\tilde{x}'(n) = \tilde{w}(n) + \mathcal{A} \tilde{s}'(n),
\end{equation}
where the SES is now given by:
\begin{eqnarray}
\nonumber
SES &=& \frac{\tilde{\textbf{x}}' \cdot \tilde{\textbf{s}}}{\sqrt{\tilde{\textbf{s}} \cdot \tilde{\textbf{s}}}} \\
     &=& \frac{ [\tilde{\textbf{w}} \cdot \tilde{\textbf{s}} + \mathcal{A} \tilde{\textbf{s}}' \cdot \tilde{\textbf{s}}] }{\sqrt{ \tilde{\textbf{s}} \cdot \tilde{\textbf{s}}}}.
\end{eqnarray}
Now Schwartz's inequality can be used to bound the inner product of the template and signal vectors:
\begin{equation}
(\tilde{\textbf{s}}' \cdot \tilde{\textbf{s}})^2 \leq (\tilde{\textbf{s}} \cdot \tilde{\textbf{s}}) (\tilde{\textbf{s}}' \cdot \tilde{\textbf{s}}'),
\end{equation}
or in terms of the associated unit vectors we have
\begin{equation}
(\hat{\tilde{\textbf{s}}}' \cdot \hat{\tilde{\textbf{s}}})^2 \leq 1 .
\end{equation}
So the dot product of the unit vectors has to be in the range $[-1,1]$.  As in \cite{Allen:2004gu}, this quantity is often referred to as the fitting factor (and is directly related to the correlation coefficient discussed above).  Following \cite{Allen:2004gu}, we can let
\begin{equation}
\hat{\tilde{\textbf{s}}}' \cdot \hat{\tilde{\textbf{s}}} = \cos\theta ,
\end{equation}
or, adding back in the dependence on the normalization:
\begin{equation}
\label{schwarz}
\tilde{\textbf{s}}' \cdot \tilde{\textbf{s}} = \cos\theta \sqrt{\tilde{\textbf{s}}' \cdot \tilde{\textbf{s}}'} \sqrt{\tilde{\textbf{s}} \cdot \tilde{\textbf{s}}},
\end{equation}
where, without loss of generality, we can restrict $\theta$ to be in the range $[0,\pi/2]$.  Now the template mismatch $\epsilon$ is given by:
\begin{equation}
\cos\theta = 1 - \epsilon.
\end{equation}

The mean and variance of the SES given in equations (\ref{expectedSES}) and (\ref{SESVar}) can now be re-computed to account for this mismatch:
\begin{equation}
{\langle z(n) \rangle} = \mathcal{A} \cos\theta \sqrt{\mathbb{D}'(n)},
\end{equation}
\begin{equation}
{\langle z^2(n) \rangle} = 1 + \mathcal{A}^2 \mathbb{D}'(n) \cos^2\theta ,
\end{equation}
where now $\mathbb{D}'$ is $\mathbb{D}$ with $\textbf{s}$ replaced with $\textbf{s}'$.  Now it easy to understand the significance of the fitting factor.  The optimal SNR is reduced by this factor when the data is filtered with a template $\textbf{s}$ that does not match the true signal $\textbf{s}'$ exactly.

This analysis can easily be extended now to the MES by considering the full set of pulses.  Summing over the pulse set $\tilde{\textbf{s}}_j$ is equivalent to simply using the full pulse train $\tilde{\textbf{s}}$ since each pulse is being windowed:
\begin{equation}
\tilde{\textbf{s}} = \sum_{j=1}^P \tilde{\textbf{s}}_j,
\end{equation}
\begin{equation}
\tilde{\textbf{s}} \cdot \tilde{\textbf{s}} = (\sum_{i=1}^P \tilde{\textbf{s}}_i) \cdot (\sum_{j=1}^P \tilde{\textbf{s}}_j) \ \delta_{ij} .
\end{equation}
where the Kronecker delta is used since the set of pulses are completely uncorrelated.  This then gives:
\begin{equation}
\tilde{\textbf{s}} \cdot \tilde{\textbf{s}} = \sum_{j=1}^P \tilde{\textbf{s}}_j \cdot \tilde{\textbf{s}}_j ,
\end{equation}
and substituting into (\ref{schwarz}) gives:
\begin{equation}
\sum_{j=1}^P \tilde{\textbf{s}}_j' \cdot \tilde{\textbf{s}}_j = \cos\theta \sqrt{\sum_{j=1}^P \tilde{\textbf{s}}_j' \cdot \tilde{\textbf{s}}_j'} \sqrt{\sum_{j=1}^P \tilde{\textbf{s}}_j \cdot \tilde{\textbf{s}}_j} .
\end{equation}
If we consider just the $j$'th temporal contribution then the total fitting factor must be divided up:
\begin{equation}
\tilde{\textbf{s}}_j' \cdot \tilde{\textbf{s}}_j = \lambda_j \cos\theta \sqrt{\sum_{j=1}^P \tilde{\textbf{s}}_j' \cdot \tilde{\textbf{s}}_j'} \sqrt{\sum_{j=1}^P \tilde{\textbf{s}}_j \cdot \tilde{\textbf{s}}_j} ,
\end{equation}
where the $\lambda_j$ are a set of $P$ real constants satisfying:
\begin{equation}
\sum_{j=1}^P \lambda_j = 1 .
\end{equation}
This shows quite simply that each temporal contribution has its own portion of the total mismatch of the pulse train and that they can all be different.  This makes sense since, even if the shape of the transits in the pulse train are not changing, a period mismatch between the model and true signal will easily make the fractional mismatch depend on transit time.

We are now in a position to re-calculate all the statistical properties of the relevant quantities as in equation (\ref{statProps}):
\begin{equation}
\begin{split}
&{\langle Z \rangle} = \mathcal{A} \mathcal{D} \cos\theta , \\
&{\langle Z^2 \rangle} = 1 + \mathcal{A}^2 \mathcal{D}^2 \cos^2\theta  , \\
&{\langle z_j \rangle} = \lambda_j \mathcal{A} \mathcal{D} \cos\theta , \\
&{\langle z_j^2 \rangle} = q_j + \lambda_j^2 \mathcal{A}^2 \mathcal{D}^2 \cos^2\theta , \\
&{\langle z_jz_k \rangle} = q_j\delta_{jk} + \lambda_j\lambda_k \mathcal{A}^2 \mathcal{D}^2 \cos^2\theta , \\
&{\langle \Delta z_j \rangle} = (\lambda_j - q_j) \mathcal{A} \mathcal{D} \cos\theta , \\
&{\langle (\Delta z_j)^2 \rangle} = q_j(1-q_j) + (\lambda_j - q_j)^2 \mathcal{A}^2 \mathcal{D}^2 \cos^2\theta , \\
\end{split}
\end{equation}
where again,
\begin{equation}
\mathcal{D} = \sqrt{ \sum_{j=1}^P \sum_{n=1}^N \tilde{s}^2_j(n) }
\end{equation}
Now the expectation value of the $\chi^2$ is given by:
\begin{eqnarray}
\nonumber
\langle \chi^2 \rangle &=& P - 1 + \kappa \mathcal{A}^2 \mathcal{D}^2 \cos^2\theta \\
                       &=& P - 1 + \kappa\langle Z \rangle^2 ,
\end{eqnarray}
where $\kappa$ is given by:
\begin{eqnarray}
\nonumber
\kappa &=& \sum_{j=1}^P (\lambda_j - q_j)^2/q_j \\
\label{keq}
       &=& -1 + \sum_{j=1}^P \lambda^2_j/q_j .
\end{eqnarray}
The variance of the $\chi^2$ is then given by:
\begin{eqnarray}
\nonumber
\sigma^2 &=& \langle (\chi^2)^2 \rangle - \langle \chi^2 \rangle^2 \\
         &=& 2(P - 1) + 4\kappa\langle Z \rangle^2 .
\end{eqnarray}
So $\kappa$ is a parameter depending on the degree of mismatch as well as the noise and is manifestly positive.  In \cite{Allen:2004gu} a proof was given that this $\chi^2$ actually has a non-central $\chi^2$ distribution with $P-1$ degrees of freedom and non-centrality parameter given by $\kappa\langle Z \rangle^2$ .  

As in \cite{Allen:2004gu}, we can obtain an upper limit on $\kappa$ from Schwarz's inequality:
\begin{eqnarray}
\nonumber
(\hat{\tilde{\textbf{s}}}_j' \cdot \hat{\tilde{\textbf{s}}}_j)^2 &\leq& (\hat{\tilde{\textbf{s}}}_j' \cdot \hat{\tilde{\textbf{s}}}_j')(\hat{\tilde{\textbf{s}}}_j \cdot \hat{\tilde{\textbf{s}}}_j) \\
\nonumber
\lambda^2_j \cos^2\theta &\leq& q_j (\hat{\tilde{\textbf{s}}}_j' \cdot \hat{\tilde{\textbf{s}}}_j') \\
\lambda^2_j/q_j &\leq& \frac{1}{\cos^2\theta}(\hat{\tilde{\textbf{s}}}_j' \cdot \hat{\tilde{\textbf{s}}}_j') .
\end{eqnarray}
Now, summing both sides over $j$ and using (\ref{keq}), we have:
\begin{equation}
0 \leq \kappa \leq \frac{1}{\cos^2\theta} - 1 .
\end{equation}

Since we have no prior knowledge of the mismatch between the true astrophysical signal and the template, we have no choice but to pick a suitable value for the purpose of estimating the non-centrality parameter.  To this end, a Monte Carlo study has been performed to get an estimate of the integral average of the signal/template mismatch.  The true astrophysical signal is generated exaclty as it is described above for DV, namely, using the Mandel-Agol geometric transit model with the non-linear limb darkening of Claret. The impact parameter has been sampled uniformly in the range $[0,1]$, the model transit durations have been sampled  uniformly in the range $[1.5,15]$, the transit depths have been sampled logarithmically (to favor weaker signals) in the range $[10^{-4.3},10^{-1.5}]$, and the stellar parameters used for limb darkening are pulled randomly from the set of $~180,000$+ targets in the Kepler Input Catalog (KIC) that are routinely searched.  I have assumed the noise is WGN. 

For this study, we focus on a single transit pulse rather than the full pulse train.  However, the potential mismatch in period has been used to expand the range on the allowable epoch mismatches, so in effect it is being taken into account.  In TPS, the allowed period mismatch is
\begin{equation}
\Delta T = 4(1-\rho) d/N ,
\end{equation}
where $\rho$ is the correlation coefficient (set to 0.9), $d$ is the transit duration, and $N$ is the number of transits. The epoch mismatch is then randomly sampled between:
\begin{equation}
-4(1-\rho) d/3 \leq \Delta t_0 \leq 4(1-\rho) d/3 ,
\end{equation}
where $d$ is the pulse duration in the set that TPS uses that most closely matches the randomly selected model value and $N$ is set to $3$ here since TPS requires a minimum of $3$ transits. The epoch and transit duration mismatches have both been turned off, simultaneously, turned on, and also tested individually.  

Two sets of pulse shapes have been used.  One set is comprised of the square wave pulses that TPS currently employs.  The other set has been generated by using this Monte Carlo framework to build a set of normalized templates, averaged over randomly selected astrophysical models taken from the parameter space of interest mentioned above.  This set of templates should on average have the best possible match to some random, true astrophysical signal.  Directly searching over the astrophysical parameter space in TPS would add too much additional volume to the search parameter space and would push us outside the computational realm of feasability with our current resources.  However, using these astrophysically motivated templates, that have been averaged over the parameter space, will be the subject of an investigation in the near future and could help us to achieve better detection efficiency and may also improve the vetoing power of the $\chi^2_{(2)}$ veto (more details in section~\ref{s:thresholds}).  The results of this study are given in Table \ref{MCTable}. 
\begin{table*}
\caption{Signal/Template Mismatch Results \label{MCTable}}
\begin{center}
\hfill{}
\small{
\begin{tabular}{cc|C{2cm}|C{2cm}|C{2cm}|C{2cm}|}
\cline{3-6} 
& & \multicolumn{2}{|c|}{Square Wave Model} & \multicolumn{2}{|c|}{Astrophysical Model} \tabularnewline
\cline{1-6} 
\multicolumn{1}{|c|}{$\Delta t_0$} & \multicolumn{1}{|c|}{$\Delta d$} & \multicolumn{1}{|C{1.5cm}|}{$\bar{\epsilon}(\%)$} & \multicolumn{1}{|C{1.5cm}|}{$\delta\bar{\epsilon}(\%)$} & \multicolumn{1}{|C{1.5cm}|}{$\bar{\epsilon}(\%)$} & \multicolumn{1}{|C{1.5cm}|}{$\delta\bar{\epsilon}(\%)$} \tabularnewline
\cline{1-6}
\multicolumn{1}{|c|}{N} & \multicolumn{1}{|c|}{N} & \multicolumn{1}{|C{1.5cm}|}{3.91} & \multicolumn{1}{|C{1.5cm}|}{0.013} & \multicolumn{1}{|C{1.5cm}|}{1.49} & \multicolumn{1}{|C{1.5cm}|}{0.0032} \tabularnewline
\cline{1-6}
\multicolumn{1}{|c|}{Y} & \multicolumn{1}{|c|}{N} & \multicolumn{1}{|C{1.5cm}|}{6.88} & \multicolumn{1}{|C{1.5cm}|}{0.021} & \multicolumn{1}{|C{1.5cm}|}{4.14} & \multicolumn{1}{|C{1.5cm}|}{0.018} \tabularnewline
\cline{1-6}
\multicolumn{1}{|c|}{N} & \multicolumn{1}{|c|}{Y} & \multicolumn{1}{|C{1.5cm}|}{4.32} & \multicolumn{1}{|C{1.5cm}|}{0.015} & \multicolumn{1}{|C{1.5cm}|}{1.66} & \multicolumn{1}{|C{1.5cm}|}{0.0051} \tabularnewline
\cline{1-6}
\multicolumn{1}{|c|}{Y} & \multicolumn{1}{|c|}{Y} & \multicolumn{1}{|C{1.5cm}|}{7.21} & \multicolumn{1}{|C{1.5cm}|}{0.019} & \multicolumn{1}{|C{1.5cm}|}{4.31} & \multicolumn{1}{|C{1.5cm}|}{0.015} \tabularnewline
\cline{1-6}
\end{tabular}
} 
\hfill{}
\tablecomments{Results for the signal/template mismatch Monte Carlo study.  Each result was generated with $50,000$ trials.  The $\delta\bar{\epsilon}$ is the error in each mean mismatch value, $\bar{\epsilon}$, given by $\sigma / \sqrt{N}$ ($N$ being the number of trials or samples).}
\end{center}
\end{table*}

From the table it is evident that the mismatch in transit duration is a small effect whereas period/epoch mismatch raises the average consistently by $~2-3\%$.  Even with the astrophysical model the mismatch for a single pulse is on average around $~4.3\%$, which is expected given that $\rho=0.9$.  This can be lowered by moving to a smaller spacing along the period dimension of parameter space at the price of increasing the computational time.  This may prove to be better than searching over the space of true astrophysical models since that would require increasing the dimensionality of the parameter space.  This will be the subject of future investigation.  It is clear from the results however, that using the set of averaged astrophysical templates cuts the mismatch down by about a factor of $2$.

\section{Thresholding Conditions}
\label{s:thresholds}

If the signal and template matched perfectly, then the results of section~\ref{s:chisquares} show that our test statistic would be $\chi^2$ distributed.  In this case, a threshold could be set by simply evaluating the $\chi^2$ cumulative distribution function.  In reality however, there is mismatch between the signal and templates, so our test statistic has a non-central $\chi^2$ distribution.  The threshold, $\chi^2_{\ast}$, that we choose then depends on the expected SNR and the mismatch.  Still, if we had a perfect understanding of the mismatch then we could set a reasonable threshold by evaluating the non-central $\chi^2$ cumulative distribution function.  Since we have no prior knowledge of the true astrophysical signal however, one way to proceed, as described in the previous section, is to understand what the integral average of the mismatch is and try and minimize it in any way possible (e.g. by using astrophysically motivated templates).  Clearly, we need to allow some room for the true spread in mismatch when we set our threshold.  The threshold for each candidate event will then be a function of the number of degrees of freedom and the non-centrality parameter.  

In \cite{Allen:2004gu}, one approach suggested is based around the fact that near the distribution maximum, when the non-centrality parameter is large compared to the degrees of freedom, the non-central $\chi^2$ distribution can be approximated by a gaussian of width $\sigma$, where $\sigma$ is the standard deviation of the non-central $\chi^2$ distribution.  So in that case, it is suggested that a reasonable threshold would be:
\begin{equation}
\chi^2_{\ast} = \langle \chi^2 \rangle + p \sigma ,
\end{equation}
where $p$ is a parameter that can be tuned empirically but would typically be something like ~$5$.  This sort of threshold has been investigated for use in TPS and is a subject of continued work.  To date however, this method has not been successful, largely it is thought due to the variance of the set of true mismatches.  So using this sort of thresholding scheme may hinge on our ability to lower the signal/template mismatch as described previously.  

An alternative thresholding method that seems to work well has additionally been investigated and is mentioned in \cite{Allen:2004gu} and described for TPS in \cite{Tenenbaum:2012ft}. For completeness, it is now reproduced here.  The dependence of the mean of $\chi^2_{(2)}$ on the degrees of freedom can be eliminated by dividing it out and forming what is commonly referred to as the reduced $\chi^2$, or, $\chi^2_r$ (assuming the signal and template matches perfectly and we have a $\chi^2$ distribution rather than the non-central $\chi^2$ discussed above):
\begin{equation}
\chi^2_r = \frac{\chi^2}{P-1}.
\end{equation}
After doing this the expectation value becomes unity
\begin{equation}
{\langle \chi^2_r \rangle} = 1.
\end{equation}
Now we could simply choose a value of $\chi^2_r$ to threhold on empirically by testing it on prior search results.  While doing this testing however it was noticed that there is some small advantage to thresholding on a different quantity that essentially converts the quantity into units equivalent to $\sigma$.  The study was done using some of the earliest $\chi^2$ results from the pipeline and has since been revisited now that all the subtleties pointed out in section~\ref{s:chisquares} have been addressed.  The quantity we currently threshold on in the pipeline is given by:
\begin{eqnarray}
\nonumber
\eta_\ast &=& \frac{MES}{\sqrt{\chi^2_r}} \\
         &=& \frac{MES}{\chi_r} .
\end{eqnarray}
We currently use a value of $7.0$ for $\eta_\ast$ (note that the MES threshold is set to $7.1$ based on a performance study in \cite{Jenkins}).  We currently use both $\chi^2_{(1)}$ and $\chi^2_{(2)}$ for vetoing purposes in this manner.  Since the $\chi^2_{(1)}$ formalism is flawed by the fact that there is overlap in the wavelet components, we are working to replace that version with $\chi^2_{(3)}$.  For this reason, examples below will only be given for $\chi^2_{(2)}$.

\section{Examples}
\label{s:examples}

\begin{figure}
\begin{center}
\epsfig{file=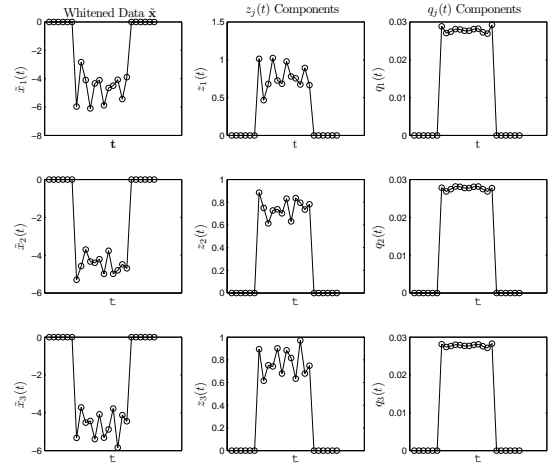,width=3.4in}
\caption{ \label{f:ex1} $\tilde{\textbf{x}}$, $z_j$, and $q_j$ for three injected $5 \sigma$ square waves in Gaussian noise.  }
\end{center}
\end{figure}

Since the method of thresholding, the model pulse shape, and the versions of the $\chi^2$ statistic used for vetoing are a subject of ongoing work, the goal here is to present some basic examples that utilize the current state of the art.  Begin by injecting three square wave pulses, each with a duration of $12$ samples, equidistant from one another in a $1,024$ sample time series.  The square waves will be injected on top of zero mean, unit variance, Gaussian noise and will each have a depth of $5 \sigma$.  Figure \ref{f:ex1} shows the whitened data chunks $\tilde{\textbf{x}}_j$ and the corresponding $z_j$ and $q_j$ for each transit.  
For this example, the MES given by equation (\ref{MES}) is $19.1$ whereas the sum over all the $z_j$ is $27.76$.  The $z_j$ components and $q_j$ components, given by summing over the points in the corresponding plots, are:
\begin{eqnarray}
\nonumber
&z_1 = 9.33, \ \  q_1 = 0.334&  \\
\nonumber
&z_2 = 9.03, \ \  q_2 = 0.332&  \\
\nonumber
&z_3 = 9.32, \ \  q_3 = 0.333&  \\
&Z = z_1 + z_2 + z_3 = 27.76 \ \ .
\end{eqnarray}
Note that these values have been rounded.  The $\chi^2$ can then be computed:
\begin{equation}
\chi^2_{(2)} = \sum_{j=1}^3 (z_j - q_jZ)^2/q_j = 0.124 \ \ ,
\end{equation}
with an associated probability computed by equation (\ref{cumProb}):
\begin{equation}
P_{\chi^2 \geq 0.124} = 1 - P_{\chi^2<0.124} = 94\% \ \ ,
\end{equation}
and $\eta$ given by:
\begin{equation}
\eta = MES/\chi_r = 76.6 \ \ .
\end{equation}
So this is a very simple detection as it should be, but it serves to illustrate how the calculation works.

Now, the same three square waves will be injected except that the last one will have its duration halved by a factor of two. The results obtained are now:
\begin{eqnarray}
\nonumber
&z_1 = 9.22,  \ \ q_1 = 0.334& \\
\nonumber
&z_2 = 8.89,  \ \ q_2 = 0.332& \\
\nonumber
&z_3 = 4.43,  \ \ q_3 = 0.333& \\
&Z = z_1 + z_2 + z_3 = 22.54 \ \ .
\end{eqnarray}
The MES is then $17.05$ and the $\chi^2$, probability, and $\eta$ can then be computed:
\begin{equation}
\chi^2_{(2)} = \sum_{j=1}^3 (z_j - q_jZ)^2/q_j = 42.89 \ \ ,
\end{equation}
\begin{equation}
P_{\chi^2 \geq 42.89} = 1 - P_{\chi^2<42.89} = 4.8\times10^{-8}\% \ \ ,
\end{equation}
\begin{equation}
\eta = MES/\chi_r = 3.68 \ \ .
\end{equation}
This illustrates how sensitive the veto is to mismatch in signal and template.

As another example, the same three square waves will be injected except the last one will now have its amplitude reduced by a factor of two.  The results obtained are now:
\begin{eqnarray}
\nonumber
&z_1 = 8.80,  \ \ q_1 = 0.335&  \\
\nonumber
&z_2 = 8.50,  \ \ q_2 = 0.332&  \\
\nonumber
&z_3 = 4.30,  \ \ q_3 = 0.334&  \\
&Z = z_1 + z_2 + z_3 = 21.60 \ \ .
\end{eqnarray}
The MES is then $16.28$ and the $\chi^2$, probability, and $\eta$ can then be computed:
\begin{equation}
\chi^2_{(2)} = \sum_{j=1}^3 (z_j - q_jZ)^2/q_j = 38.20 \ \ ,
\end{equation}
\begin{equation}
P_{\chi^2 \geq 38.20} = 1 - P_{\chi^2<38.20} = 5.07\times10^{-7}\% \ \ ,
\end{equation}
\begin{equation}
\eta = MES/\chi_r = 3.73 \ \ .
\end{equation}
So the veto is sensitive to a mismatch in depth of one of the pulses as well.  

A logical extension of this would be to ask: how do these numbers change when there are more transits and still only a single transit is perturbed in some way?  When the degrees of freedom get large enough, the effect will get washed out to the point that the veto will no longer work.  This behavior is acceptable in this case however, because if there are say, $10$ transits, and only one of them is perturbed in some way, we would still have reasonable belief there is something interesting in the data.  Clearly though, if many of the transits do not match well then the veto will work as it should.    

\section{Results}
\label{s:results}
As described in \cite{Tenenbaum:2012ft}, events that pass the MES threshold of $7.1$ in TPS are then subjected to both a robust statistic test as well as the two $\chi^2$ tests mentioned in section~\ref{s:thresholds}, in that order.  Recently, TPS was run over a set of $192,312$ targets, (known eclipsing binaries were all removed), with quarters $Q1-Q12$, and with the $\chi^2$ thresholds turned down to gain an understanding of their operating characteristics.  The set of Kepler Objects of Interest (KOI's) was used to understand the detection probability.  The set of KOI's is comprised mostly of known planets as well as planet candidates which have not gone through the vetting process.  A set of $2,211$ KOI stars, the best of the best, were set aside to examine the detection probability.  To understand the false alarm probability, the full set of $3009$ unique KOI target stars were removed from the total set of targets, the remainder of which forms the false alarm population of targets.  Of course, some of these false alarms may actually be true positives, but it is thought that the number is fairly low.  It should also be noted that the false alarm and detection probabilities gleaned by a study of this sort are only to be used as an indicator since the true pipeline version of TPS allows for searching over many period and epoch combinations, for each target, that produce a MES above threshold.  Here we have limited the number of searched combinations to only one per target.  

\begin{figure}
\begin{center}
\epsfig{file=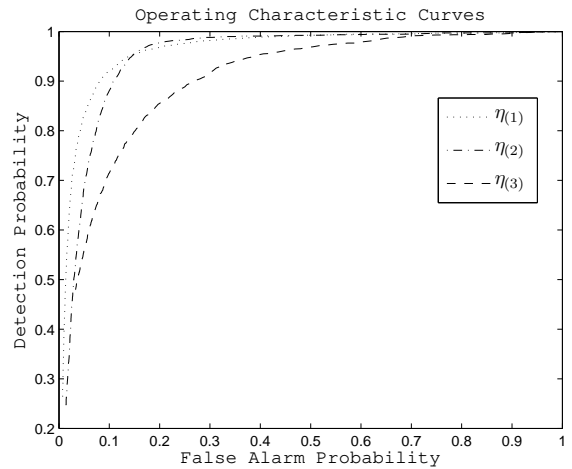,width=3.4in}
\caption{ \label{f:OC} Operating characteristic curve for the three vetoes of interest.  }
\end{center}
\end{figure}

\begin{figure}
\begin{center}
\epsfig{file=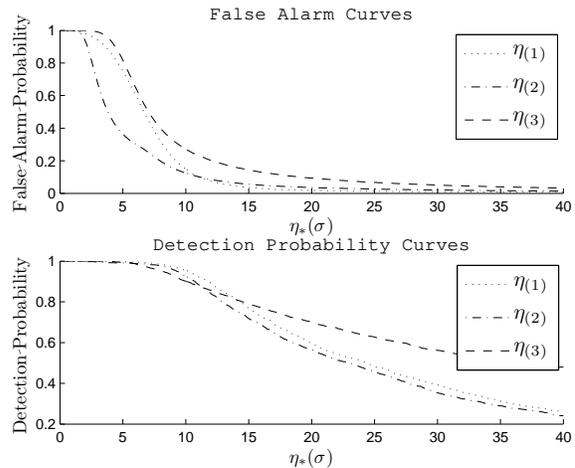,width=3.4in}
\caption{ \label{f:probs} False alarm and detection probabilities corresponding to the operating characteristic curves in figure \ref{f:OC}.  }
\end{center}
\end{figure}

After applying the MES threshold of $7.1$, there are $14$ targets that get filtered out of the $2,211$ KOI's, giving a $99.4\%$ detection probability.  There are then $86,737$ non-KOI targets that produce ``false alarms'', for a false alarm rate of $45.8\%$.  Applying the robust statistic threshold of $\eta = 6.4$ filters out an additional $13$ of the $2,211$ KOI's for an overall detection probability of $98.8\%$.  The number of non-KOI targets producing false alarms is reduced to $40,880$, for an overall false alarm rate of $21.6\%$.  Now we will just use these two sets of remaining targets namely, the set of $2,184$ KOI's and the set of $40,880$ non-KOI's, to explore various thresholds on $\eta_{(m)}=MES/\chi_{(m)r}$.  Figure \ref{f:OC} shows the operating characteristic curve for each $\eta$ individually.  You can clearly see that $\eta_{(2)}$ achieves the best detection efficiency but would still, for some reasonable detection probability, let through an unacceptable number of false alarms if used alone. In figure \ref{f:probs}, the corresponding false alarm and detection probabilities are plotted as a function of threshold $\eta_{\ast}$.  

As mentioned in \cite{Tenenbaum:2012ft}, TPS currently uses $\eta_{(1)}$ and $\eta_{(2)}$ for vetoing purposes, the threshold for both set to $7.0$.  Using $\eta_{(3)}$ instead of $\eta_{(1)}$ is the subject of ongoing work but is desirable due to the fact that the $\chi^2_{(1)}$ formalism is flawed for reasons mentioned previously.  On this data set, using the current TPS thresholds gives an overall (MES, robust statistic, $\eta_{(1)}$, and $\eta_{(2)}$ all being used) detection probability of $~97\%$ and an overall false alarm probability of $~3.77\%$ or $7,133$ false alarm targets.   If we were to replace $\eta_{(1)}$ with $\eta_{(3)}$, and require the same detection probability, then the threshold for $\eta_{(3)}$ would have to be set at $4.8$ and would give an overall false alarm probability of $~4.89\%$, or $9,260$ targets.  Further investigation reveals that there is an issue with the fitted depth $\mathcal{A}$ being used in the calculation of $\eta_{(3)}$.  The fitted depth comes from the robust fit of a whitened model pulse train to the whitened data when the robust statistic is being calculated.  This value can be perturbed by the robust statistic algorithmn however, since it has some machinery built in that allows it to deemphasize cadences based on the robust fit weights.  This can throw off the comparison between the observed and expected values.  To address this problem, and hopefully improve the detection efficiency of $\eta_{(3)}$ enough to make it a more suitable replacement, the fitted depth is now being calculated by:
\begin{equation}
\mathcal{A} = \frac{\tilde{\textbf{x}} \cdot \tilde{\textbf{s}}_j}{\tilde{\textbf{s}}_j \cdot \tilde{\textbf{s}}_j}  . \\
\end{equation} 
We await future results.

Another recent run of TPS was done over the same set of quarters and $192,255$ targets.  In this run however, the full looping machinery was employed so that TPS was examining up to $1,000$ events for each target that produced a sufficiently high MES.  In this set of targets there were $2,264$ targets with high quality KOI's, and $3,008$ total KOI's in the sample.  There were $13,570$ targets producing Threshld Crossing Events (TCEs).  Our detection probability in this run was $~96\%$ whereas the overall false alarm probability was $~5.6\%$.  In this run, using $\eta_{(1)}$ and $\eta_{(2)}$ with thresholds both at $7.0$ caused $~2\%$, or about half, of the total loss in detection probability.  Using the vetoes however dropped the number of false alarms from $55,233$ down to $10,694$, or from $~29\%$ overall false alarm rate down to just $~5.6\%$.  

\clearpage

\section{Conclusion}
\label{s:conclusion}
This paper extends the conceptual framework in \cite{Allen:2004gu} so that the formalism can be applied to the case of interest here, namely, in vetoing false alarms coming from the Transiting Planet Search component of the \textit{Kepler} data processing pipeline.  We have presented a set of potential vetoes and selected two, $\chi^2_{(2)}$ and $\chi^2_{(3)}$, to be the subject of future work and development based upon the results presented here.  These vetoes cut down the false alarm rate considerably and are crucial in maintaining a large enough search parameter space and detection probability.  Further mitigation of some of the known sources of systematic error (image artifacts, etc), combined with improvements to this set of vetoes (using better templates and switching to the method of thresholding described in section~\ref{s:thresholds}) are other avenues being pursued to improve detection efficiency.

\acknowledgments

The author wishes to thank Bruce Allen for the original work on this subject from which most of this work is based.  \textit{Kepler} was selected as the $10^{th}$ mission of NASA's Discovery Program.  Funding for this work is provided by NASA's Science Mission Directorate.

\clearpage

\end{document}